\documentclass[a4paper,notitlepage]{article}

\usepackage[english]{babel}
\usepackage[utf8]{inputenc}
\usepackage[T1]{fontenc}
\usepackage{amssymb}
\usepackage{braket}
\usepackage{bm}
\usepackage{authblk}

\usepackage[a4paper,top=3cm,bottom=3cm,left=2.9cm,right=2.9cm,marginparwidth=1.75cm]{geometry}

\usepackage{amsmath}
\usepackage{graphicx}
\usepackage[colorinlistoftodos]{todonotes}
\usepackage[colorlinks=true, allcolors=blue]{hyperref}
\usepackage{todonotes}
\usepackage{color}

\usepackage{caption}
\usepackage{authblk}
\usepackage{etoolbox}
\usepackage{lmodern}

\title{Adiabatic Quantum Kitchen Sinks for Learning Kernels \\Using Randomized Features}

\author[1]{Moslem Noori}
\author[1,2]{Seyed Shakib Vedaie}
\author[1]{Inderpreet Singh}
\author[1]{Daniel Crawford}
\author[1,3]{\newline Jaspreet S. Oberoi}
\author[2]{Barry C.\ Sanders}
\author[1]{Ehsan Zahedinejad}

\affil[1 ]{\small 1QB Information Technologies (1QBit), Vancouver, BC, Canada}
\affil[2 ]{\small Institute for Quantum Science and Technology, University of Calgary, Calgary, AB, Canada}
\affil[3 ]{\small School of Engineering Science, Simon Fraser University, Burnaby, BC, Canada}

\begin{document}
\maketitle

\begin{abstract}
    Quantum information processing is likely to have far-reaching impact in the field of artificial intelligence. While the race to build an error-corrected quantum computer is ongoing, noisy, intermediate-scale quantum (NISQ) devices provide an immediate platform for exploring a possible quantum advantage through hybrid quantum--classical machine learning algorithms. One example of such a hybrid algorithm is ``quantum kitchen sinks”, which builds upon the classical algorithm known as ``random kitchen sinks'' to leverage a gate model quantum computer for machine learning applications. We propose an alternative algorithm called ``adiabatic quantum kitchen sinks'', which employs an adiabatic quantum device to transform data features into new features in a non-linear manner, which can then be employed by classical machine learning algorithms. We present the effectiveness of our algorithm for performing binary classification on both a synthetic dataset and a real-world dataset. In terms of classification accuracy, our algorithm significantly enhances the performance of a classical linear classifier on the studied binary classification tasks and can potentially be implemented on a current adiabatic quantum device to solve practical problems.
\end{abstract}

\section{Introduction}
Quantum algorithms~\cite{Mon16} are theoretically proven to solve certain computational problems faster than the best known classical algorithms~\cite{Sho97,Gro96}.~Despite the impressive progress made toward building a universal quantum computer in the quest for quantum supremacy~\cite{HM17,BSL+16}, it remains an elusive goal due to the negative effects of noise present in quantum systems.~Meanwhile, noisy, intermediate-scale quantum (NISQ)~\cite{Pre18} devices readily provide a platform for demonstrating a potential quantum advantage for specific applications such as machine learning~\cite{BWP+17}.

\begin{figure}[h]
    \centering
    \includegraphics[width=0.6\textwidth]{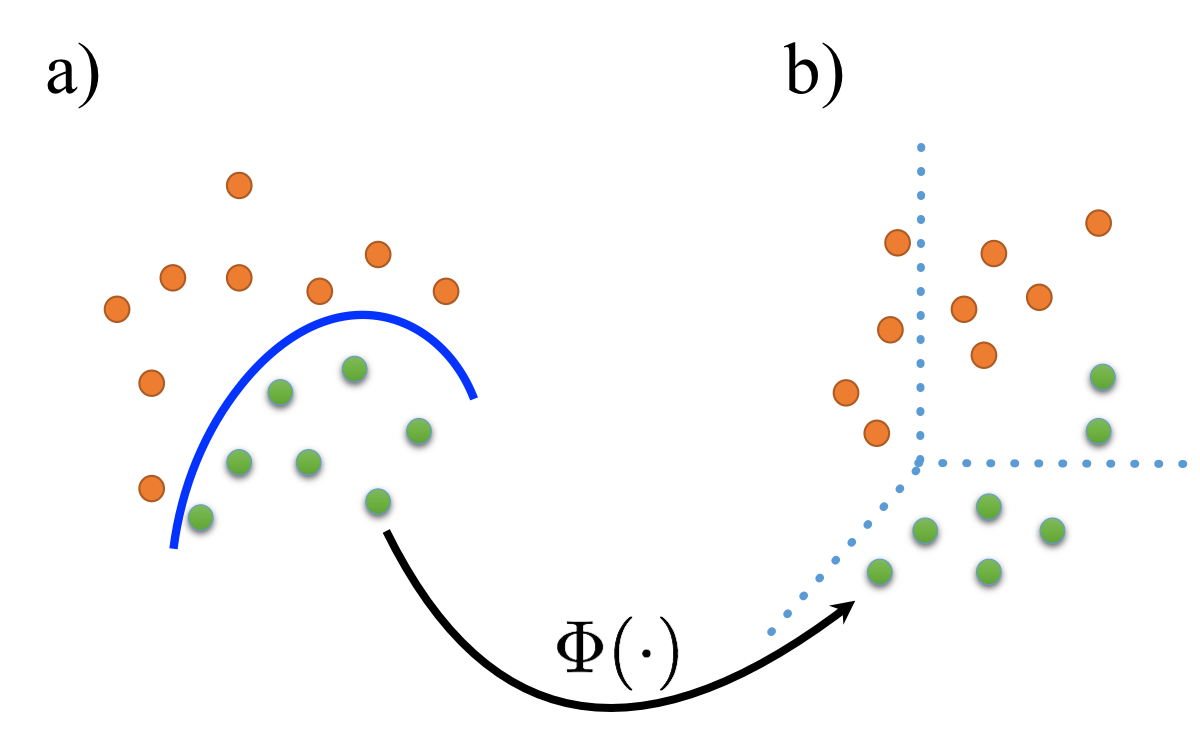}
    \caption{A classification problem where the goal is to train a classifier to distinguish two classes of orange- and green-coloured data samples. A feature map is shown that transforms a) each data sample of the two non-linearly separable classes into b) a data sample with new features in a higher-dimensional space where the two classes become linearly separable. The kernel corresponding to such a feature map is a non-linear kernel.}
    \label{fig:kernel}
\end{figure}

The main goal in machine learning is to discover patterns and learn from data.~The emergence of new classical hardware has enabled faster learning to occur using enormous quantities of data~\cite{RDG+16,KXS+16}. However, the rapid growth in the amount of available data requires increasingly faster computing devices to learn from big data, and quantum computers are a potential candidate.~Several recent studies have shown the potential of NISQ technologies in machine learning.~For example,~\cite{HCT+19} proposes a kernel-based supervised quantum machine learning (QML) algorithm that has the potential to show quantum supremacy in machine learning.~Other examples of QML algorithms using NISQ technologies are studies using quantum Boltzmann machines~\cite{AAE+18, CLG+18, LCG+17} and quantum clustering algorithms~\cite{OMA+17}. 
Kernel machines play an important role in machine learning. A kernel, $\mathcal{K}$, is a positive semidefinite matrix whose element, $\mathcal{K}({\bm{x}_i},{\bm{x}_j})$, denotes a similarity measure between a pair of data samples. Here, ${\bm{x}_i}\in\mathcal{X}$ is a $p$-dimensional vector defined over ${\mathbb{R}}^p$, $\mathcal{X}$ is the set of input data that contain $n$ data samples, and $i\in\{1,2,\ldots, n\}$. Mathematically, a kernel is defined as $\mathcal{K}({\bm{x}_i},{{\bm{x}}_j})=\braket{\Phi{(\bm{x}_i)},\Phi{(\bm{x}_j)}}$, where $\Phi:{\mathbb{R}^p}\xrightarrow{}{\mathcal{H}}$ is called an explicit feature map which transforms a given data sample into one residing in the Hilbert space ${\mathcal{H}}$, usually called the ``feature space''. The choice of kernel, for example,~linear or non-linear, dictates the performance of the underlying kernel machine (see Fig.~\ref{fig:kernel}).

Recently, an algorithm called ``quantum kitchen sinks'' (QKS) was proposed~\cite{WOT+18}, which builds upon the idea of classical ``random kitchen sinks'' (RKS)~\cite{RB08_1,RB08_2,RB09}.~It uses a gate model quantum computer as an explicit feature map to generate randomized features from the original input features. Once done, it enables a classical (linear) machine learning algorithm acting on the randomized features in the feature space $\mathcal{H}$ to be more effective in learning than it was in the original space.

In this work, we propose an alternative approach to QKS~\cite{RB08_1,RB08_2,RB09} called  ``adiabatic quantum kitchen sinks'' (AQKS). Our algorithm uses an adiabatic quantum annealer as an explicit feature map to transform the features of each data sample into new features called quantum randomized features. In short, given a data sample $\bm{x}$, we encode its input data features into the parameters of a quantum Hamiltonian. Evolving the quantum system and performing a measurement at the end of the evolution gives us a new data sample that represents $\bm{x}$ in the feature space $\mathcal{H}$. The  kernel that results from such a transformation is non-linear because of the effect of the measurement operator on the quantum system. We show that such a non-linear explicit feature map has a positive impact on learning kernel machines for classification problems.

Similar to~\cite{WOT+18}, our work can be seen as a feature engineering technique that leverages a quantum device to generate new features for classical machine learning algorithms. Unlike variational-based QML algorithms~\cite{HCT+19,MNK18,SK19}, our algorithm does not require an iterative call to a quantum device. The AQKS algorithm could provide complex non-linear transformations that have not been previously identified in classical kernel machines.

We consider the following scenario to show the effectiveness of AQKS on a learning task. Given a dataset $\mathcal{D}\subset{\mathbb{R}^p}$, we first use AQKS to construct a new dataset $\mathcal{D}'\subset\mathcal{H}$. We then train two support vector machines with a linear kernel (LSVM), one on the dataset $\mathcal{D}$ and the other on the dataset $\mathcal{D}'$. We call the first model that is trained on $\mathcal{D}$ an LSVM, and the other trained on $\mathcal{D}'$ an AQKS+LSVM model. Keeping the learning algorithm in both models the same (i.e.,~an~LSVM), we compare the performance (i.e., the classification accuracy) of the two models on two example datasets.

To demonstrate the power of AQKS for machine learning, we evaluate the performance of our algorithm on a synthetic dataset as well as on the Modified National Institute of Standards and Technology (MNIST) dataset. Our experiments show that AQKS significantly outperforms (in terms of the classification accuracy) the LSVM for classification tasks for the studied datasets. Specifically, our algorithm increases the classification accuracy of an LSVM on the synthetic dataset from $50\%$ to $99.4\%$. On the MNIST dataset, our algorithm reduces the classification error of an LSVM from $4.4\%$ to $1.6\%$. It is important to mention that the AQKS algorithm can readily be applied to practical datasets with any number of features using a current quantum annealer.

This work is structured as follows. In Section~\ref{sec:AQC}, we give a short overview of adiabatic quantum computation. Section~\ref{sec:ARKS} explains the idea behind the RKS algorithm and discusses how RKS can be connected to a adiabatic quantum device to devise a hybrid quantum--classical machine learning algorithm. Section~\ref{sec:hamiltonian_to_kernel} explains how the measurements obtained from an adiabatic quantum system relate to a non-linear kernel, which represents the effect of the AQKS algorithm on data feature engineering.~In Section~\ref{sec:experiment} we outline our experimental settings, and in Section~\ref{sec:results} we report the results of the experiments. We discuss the results in Section~\ref{sec:discussion}. Section~\ref{sec:conclusion} concludes our work and suggests directions for future research.

\section{Adiabatic Quantum Computation}
\label{sec:AQC}
Adiabatic quantum computation (AQC), proposed by Farhi et al.~\cite{Farhi2000, Farhi2001}, is a model for solving computational problems (e.g.,~combinatorial optimization problems) by slowly evolving a quantum system's Hamiltonian from an initial Hamiltonian $\pmb{\text{H}}_{\text{i}}$ to a final Hamiltonian $\pmb{\text{H}}_{\text{f}}$, which encodes the computational problem at hand. One can write the total $q$-body Hamiltonian of the system, denoted by $\pmb{\text{H}}(t)$, as
\begin{equation}
\label{eq:quantum_annealing}
\pmb{\text{H}}(t)=a(t)\pmb{\text{H}}_{\text{i}} + b(t)\pmb{\text{H}}_{\text{f}}\,,    
\end{equation}
where $a(t)$ and $b(t)$ are two time-dependent and smooth functions, which are  monotonically decreasing and increasing, respectively. By varying these functions, we evolve the system's Hamiltonian over the time interval [$0$,~$T$].~Whereas the ground state of the initial Hamiltonian is known and easy to prepare, the ground state of the final Hamiltonian is not known.~Farhi et~al.~\cite{Farhi2000} showed that if the evolution time ($T$) is sufficiently large with respect to the energy gap of the evolving quantum system, then the adiabatic theorem guarantees that at the end of the evolution, one will find the quantum system at the ground state of $\pmb{\text{H}}_{\text{f}}$ with high probability. The ground state of the final Hamiltonian represents the solution to the encoded computational problem.

Adiabatic quantum computing theory assumes that the quantum system under evolution is isolated from the surrounding environment.~Under such an assumption, AQC is polynomially equivalent to standard gate model quantum computing~\cite{Aharonov2007}.~For the case of an open quantum system with nonzero temperature, such a polynomial equivalency relationship is yet to be established.

One example of a nonzero-temperature, non-universal type of AQC is a D-Wave Systems quantum annealer, designed to implement the quantum annealing Hamiltonian (\ref{eq:quantum_annealing}) with $\pmb{\text{H}}_{\text{i}}=-\sum_{v=1}^{q} \sigma_{v}^{x}$ and \mbox{$\pmb{\text{H}}_{\text{f}}= \sum_{\braket{l,m}} h_{lm}\sigma_{l}^{z}\sigma_{m}^{z} + \sum_{u} j_{u}\sigma_{u}^{z}$}, where $\braket{l,m}$ goes over pair-wise interacting qubits and $\sigma^z$ and $\sigma^x$ are Pauli-Z and Pauli-X operators, respectively.

The primary interest in our study is to employ an adiabatic quantum computing device to perform machine learning tasks.~In the section that follows, we explain how we connect an adiabatic quantum device to machine learning algorithms.

\section{Adiabatic Quantum Kitchen Sinks}
\label{sec:ARKS}
Kernel methods are at the heart of machine learning.~Despite their impressive performance in  machine learning tasks~\cite{HSS08}, kernel methods become increasingly intractable when the applications involve big data.
To overcome these computational challenges, Rahimi et al.~\cite{RB08_1,RB08_2,RB09} proposed RKS, which involves mapping the input data samples into a randomized feature space, such that the overlap (inner product) of the pair of data samples in the randomized space approximates a desired kernel.~A linear machine learning algorithm then acts on the randomized samples generated from the input data to perform the learning process.~Despite its simplicity, the performance of RKS is comparable to state-of-the-art machine learning algorithms~\cite{RR08_4,MGL+19}.

We propose a quantum--classical hybrid machine learning algorithm that uses RKS in combination with an adiabatic quantum device. In our AQKS algorithm, we encode the features of the data samples into the parameters of an adiabatic quantum system and then evolve the system, and sample from its final state to generate the quantum randomized features. A linear classical machine learning algorithm can then be applied on the generated quantum randomized features to discover potential patterns in the dataset.

To provide a more formal description of the AQKS algorithm, let us consider the dataset $\mathcal{D}$.~The first step in transforming the given dataset from the original space (i.e.,~$\mathbb{R}^p$) to a new space (i.e.,~$\mathcal{H}$) is ``encoding''.~To this end, we define $\pmb{\text{A}}$ and ${\bm{b}}$, where $\pmb{\text{A}}$ is a $q\times{p}$ random matrix with a classical probability distribution function $P(\pmb{\text{A}})$ and ${\bm{b}}$ is a $q$-dimensional vector with a classical probability distribution function $P({\bm{b}})$. We encode ${\bm{x}}_i$ into ${\bm{j}}_i$ by applying the linear transformation
\begin{equation}
\label{eq:linear_map}
{\pmb{\text{j}}}_i = {\pmb{\text{A}}}{\bm{x}}_i+{{\bm{b}}}.
\end{equation}
The process of choosing a certain combination of $\pmb{\text{A}}$ and ${\bm{b}}$ is repeated multiple times, and we call each repetition an ``episode'', denoted by $e$.~The corresponding encoding for each episode is represented by ${\bm{j}}_i^e$.~Once the encoding process is complete, the resultant $q$-dimensional ${\bm{j}}_i^e$ is mapped onto the coefficients of the local $\sigma^z$ terms of a $q$-body transverse-field Hamiltonian, $\pmb{\text{H}}(t)_{\bm{{x}_i}}^e$, such that
\begin{equation}
\label{eq:Isingencoding}
\pmb{\text{H}}(t)^e_{\bm{x}_i}=a(t)\sum_v^{q}\sigma_v^x+\sum_{\braket{l,m}}h_{lm}\sigma^z_{l}\sigma^z_{m} + \sum_{u}^{q} j_u^e\sigma^z_u.
\end{equation}
In (\ref{eq:Isingencoding}), $j_u^e$ is the $u$-th element of the vector ${{\bm{j}}}_i^e$.~Each $h_{lm}$ is a real number derived from a function of $j_u^e$ (in our experiments we consider $h_{l,m}^e =  j_{l}^e j_{m}^e$) that could result in quantum entanglement, being the coefficient for the $\sigma_z\sigma_z$ interaction.

In order to generate a transformed data sample from each $\bm{{x}}_i$, we evolve $(\ref{eq:Isingencoding})$ on an adiabatic quantum device from an initial time $t_\text{i}=0$ to a final time $t_\text{f}=T$ and then perform a projective measurement along the z-axis at the end of the evolution.~Stacking the outcomes of the measurements generated through a total of $E$ episodes for $\bm{x}_i$ and normalizing the resultant vector by $\frac{1}{E}$ provides a ($q\times{E}$)-dimensional vector ${\bm{u}}_{\bm{x}_i}$, which represents the respective $\bm{x}_i$ in the feature space.

In summary, AQKS comprises three steps. First, we encode the data into the parameters of a quantum Hamiltonian using the encoding formula~\eqref{eq:linear_map}. Second, we evolve the Hamiltonian of the quantum system for a time duration $T$. Finally, we collect the quantum randomized features through measurement. Note that the first step is the linear transformation of a data sample $\bm{x}$ into a new vector ${\bm{j}}$. Therefore, when we assign the elements of ${\bm{j}}$ into the parameters of the Hamiltonian, we are not introducing any non-linearity into the quantum system's Hamiltonian. This ensures that if any non-linear behaviour is observed from AQKS, it can be completely attributed to the quantum device and not the manner in which we encode data into the quantum system's Hamiltonian.

\section{From Quantum Hamiltonian to Non-linear Kernel}
\label{sec:hamiltonian_to_kernel}
In this section, we relate AQKS to kernel methods.~We follow the same approach as explained in~\cite{WOT+18} and modify the specifics as needed to connect the random kitchen sinks theory to the adiabatic quantum device.~A kernel is a mathematical object that defines a similarity metric between any two samples in a Hilbert space.~For example, we can define the inner product between two data samples as a kernel $\mathcal{K}(\bm{x}_n, \bm{x}_m)$ that can be represented as
\begin{equation}
\label{eq:kernel}
\mathcal{K}(\bm{x}_n, \bm{x}_m)=\braket{\bm{x}_n,\bm{x}_m}.
\end{equation}
AQKS is concerned with achieving a non-linear transformation of the data via an adiabatic quantum device, that is,~mapping the dataset into a feature space $\mathcal{H}$, similar to what an explicit feature map does in the classical case.~Note that our algorithm does not explicitly calculate the kernel, but implicitly produces a similar effect on the data.~We now discuss the type of non-linearity that a quantum annealing device (or, more generally, an adiabatic quantum device) can generate through the data transformation procedure we introduce in the previous section.

Given the quantum system's Hamiltonian~\eqref{eq:quantum_annealing}, corresponding to the encoding of a data sample ${\bm{x}_m}$, the unitary evolution of the adiabatic quantum device is
\begin{equation}
\label{eq:evolution}
\pmb{\text{U}}(\bm{x}_m,e,T) = \mathcal{T}\text{exp}\left(-\text{i}\int_0^{T}\pmb{\text{H}}(t)^e_{\bm{x}_m}\text{d}t\right),
\end{equation}
    where $\mathcal{T}$ is the time-ordered operator and all of the other symbols have their usual meaning. The~unitary operator~\eqref{eq:evolution} evolves the quantum system from an initial state $\ket{\psi}_\text{i}$ to a final state $\ket{\psi_{\bm{x}_m}}_\text{f}=\pmb{\text{U}}(\bm{x}_m,e,T)\ket{\psi}_\text{i}$. By performing a measurement on $\ket{\psi_{\bm{x}_m}}_\text{f}$ at each episode and concatenating $E$ binary vectors of length $q$, we form a binary vector $\bm{u}_{\bm{x}_m}$. Note that the outcome of the measurement at each episode is a random binary vector $\bm{z}\in\{0,1\}^q$ with a probability $p^e_{\bm{x}_m,\bm{z}}$ given by
\begin{equation}
\label{eq:probability}
p^e_{\bm{x}_m,\bm{z}} = |\braket{\bm{z}|\pmb{\text{U}}(\bm{x}_m,e,T)|\psi_\text{i}}|^2.
\end{equation}

Now, let us consider $\bm{u}_{\bm{x}_m}$ and $\bm{u}_{\bm{x}_n}$ as two binary vectors in the feature space that correspond to two data samples $\bm{x}_m$ and $\bm{x}_n$, respectively. Using the kernel definition~\eqref{eq:kernel}, we define the element of the quantum kernel, ${\mathcal{K}}(\bm{x}_m,\bm{x}_n)$, to be a quantity proportional to the inner product of the two data samples $\bm{u}_{\bm{x}_m}$ and $\bm{u}_{\bm{x}_n}$. Mathematically, we express the quantum kernel as 
\begin{equation}
\label{eq:kernel_transformed}
\mathcal{K}(\bm{x}_m,\bm{x}_n)= \frac{1}{E}\braket{\bm{u}_{\bm{x}_m},\bm{u}_{\bm{x}_n}}.
\end{equation}
We rewrite the right-hand side of (\ref{eq:kernel_transformed}) in the form
\begin{equation}
\label{eq:compact_form}
\braket{{\bm{u}}_{\bm{x}_m},{\bm{u}}_{\bm{x}_n}} =\frac{1}{E}\sum_{e=1}^E \braket{{\bm{p}}^e_{\bm{{x}}_m},\pmb{\text{S}}{\bm{p}}^e_{\bm{{x}}_n}},
\end{equation}
where each element $\text{s}_{i,j}(i,j\in\{1,\ldots, q\})$ of matrix $\pmb{\text{S}}$ is defined as $\text{s}_{i,j}=\bm{d}_i\bm{d}_j^T$ and $\bm{d}_i$ is the corresponding $q$-dimensional binary vector representing the integer $i$. Here, ${\bm{p}}^e_{\bm{{x}}_m}$ is a $q$-dimensional vector where each of its element is given by \eqref{eq:probability} representing the outcome probabilities from measurement at each episode $e$.

The right-hand side of (\ref{eq:compact_form}) can be realized as a Monte Carlo estimate of the kernel that converges to the true value of the kernel $\mathcal{K}(\bm{{x}}_m, \bm{{x}}_n)$ as the number of episodes approaches infinity.~We thus write the kernel
\begin{equation}
\label{eq:quantum_kernel_int}
{\mathcal{K}}(\bm{{x}}_m, \bm{{x}}_n)= \int\hspace{-0.15cm}\int{{P({\pmb {\text{A}}}) P({{\bm{b}}}) \, {\bm{p}}({\pmb{\text{A}}, {\bm{b}}})^T_{\bm{{x}}_m}} \, \pmb{\text{S}} \, {\bm{p}}({\pmb{\text{A}}, \bm{{b}}})_{\bm{{x}}_n}} \, \text{d}{\pmb{\text{A}}} \, \text{d}{{\bm{b}}}\,,
\end{equation}
wherein~\eqref{eq:quantum_kernel_int} we have explicitly shown the dependency on ${\bm{p}}_{\bm{{x}}_m}$ to ${\pmb{\text{A}}}$ and ${{\bm{b}}}$.~A methodological measurement on an adiabatic quantum device results in the non-linear transformation explained above. Whereas it is straightforward to explicitly express the mathematical formula of a quantum kernel corresponding to a two-qubit quantum circuit (see, e.g.,~\cite{WOT+18}), it is not trivial to do so in the case of a quantum kernel obtained using adiabatic quantum computing~\eqref{eq:quantum_kernel_int}. This is because driving an analytical expression for the elements of ${\bm{p}}_{\bm{{x}}_m}$~becomes challenging in the case of an adiabatic quantum process.

\section{Experimental Setting}
\label{sec:experiment}
In this section we discuss the experimental settings used to evaluate the classification performance of our algorithm, AQKS.

\subsection{Performance Measure}
After applying quantum randomization on input data samples, data is mapped into a higher-dimensional space where we expect a relatively simple classifier, for example,~an LSVM, to classify the data effectively.~To measure the level of success in achieving this goal, we first create a baseline for the underlying learning task by solving the classification problem using an LSVM without quantum randomization (i.e.,~with the data residing in its original space).~We then compare the performance of the AQKS against the mentioned baseline. In other words, we apply an LSVM after applying quantum randomization on the original data and compare its performance against an LSVM when it is used without quantum randomization. This provides a systematic way to fairly assess the power of a quantum device as a non-linear explicit feature map.

\subsection{Datasets}
\begin{figure}[h]
    \centering
    \includegraphics[width=0.5\textwidth]{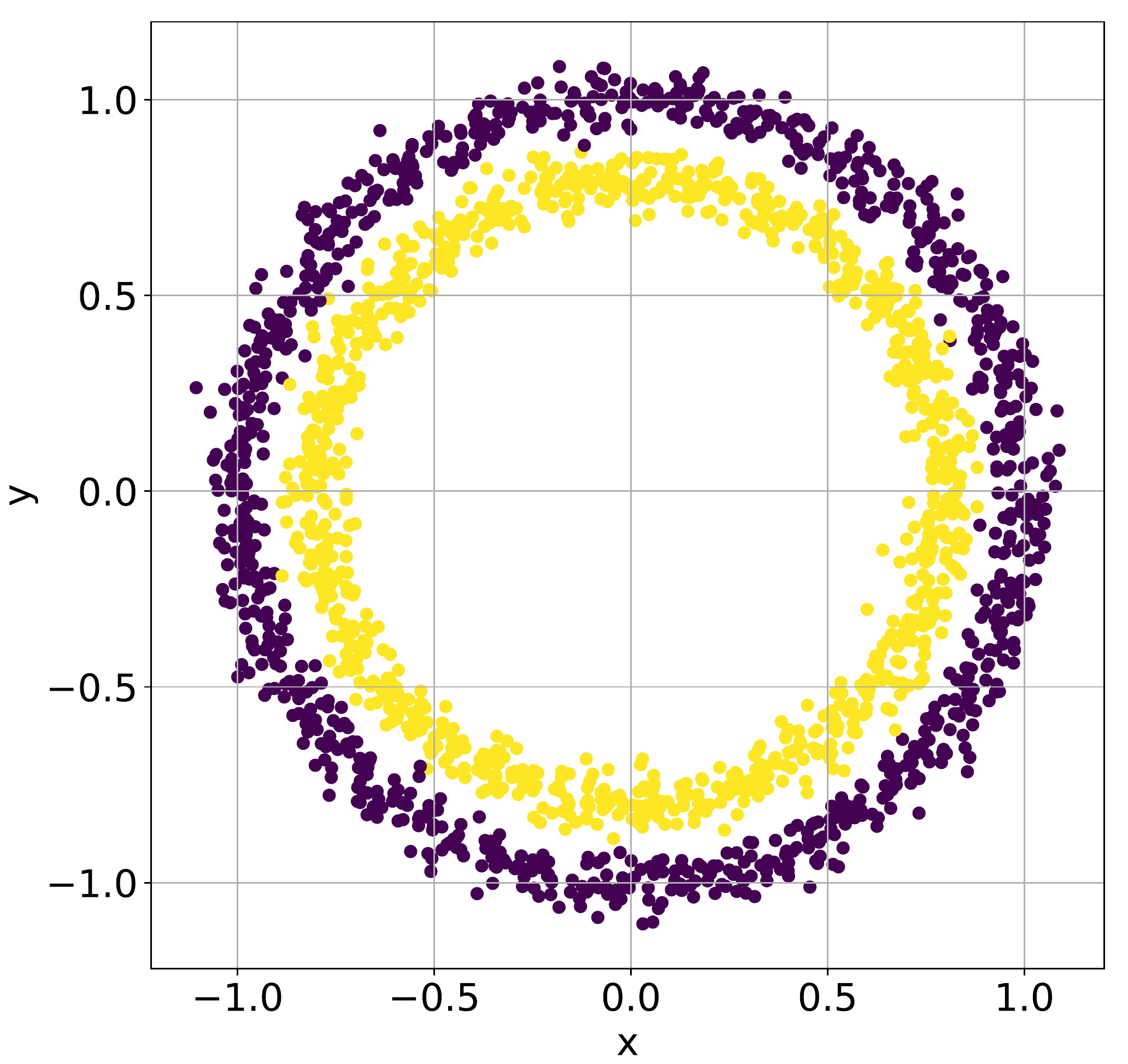}
    \caption{Representation of the “circles” dataset generated using the “datasets” module from \mbox{“scikit-learn”}. Two classes (yellow and purple), based on an arbitrary radius from the datasets module and consisting of 1000 data samples each, are shown (see the main text for more details on the setting of the generated dataset).}
    \label{fig:circle data}
\end{figure}

We evaluate the performance of our algorithm on two datasets.~The first  is a two-dimensional synthetic dataset, consisting of two classes, and generated using the {\tt sklearn.datasets} Python module (see Fig.~\ref{fig:circle data}).~The ratio of the inner circle (class 1) radius to that of the outer circle \mbox{(class 2)} is $0.8$, and the standard deviation of the Gaussian noise added to the data  is $0.04$.~This dataset is linearly inseparable in the two-dimensional space, making it a good candidate for studying the effect of quantum randomization on the classifier's accuracy.

\vspace{1.1em}

\vspace{1.1em}

The second dataset we consider is the MNIST dataset, a practical dataset widely used for testing and benchmarking machine learning algorithms.~The dataset contains a large ensemble of handwritten digits, where each data sample is a $28$-by-$28$-pixel greyscale image.~Each image can be represented by a $784$-dimensional vector ${\bm{x}}$ whose elements represent the shade, in grey, of the pixels, and ranges from 0 to 255.~We evaluate the performance of our algorithm in classifying the handwritten digits ``3'' and ``5''.~The dataset contains 7141 and 6313 instances of the digits 3 and 5, respectively.

\subsection{Simulating the Adiabatic Quantum Evolution}
We Trotterize in order to simulate the quantum system evolution~\cite{WBL02}. The evolution of the quantum system's Hamiltonian $\pmb{\text{H}}(t)$ over the time span $[0,T]$ is decomposed into short time steps during which the quantum Hamiltonian is approximately time-independent.~As a result, the approximated evolution operator, $\pmb{\tilde{\text{U}}}(T)$, of $\pmb{\text{H}}(t)$, is
\begin{equation} \label{eq: trotterization}
    \pmb{\tilde{\text{U}}}(T) \approx \prod_{a=0}^{k-1}\exp[-\text{i}\pmb{\text{H}}(a\tau) \tau]+\epsilon,
\end{equation}
where $k$ is the number of time steps, $\tau = \frac{T}{k}$ is the duration of each time step, and $\epsilon$ subsumes terms of order $\tau^2$ and higher. 

\subsection{The Effect of the Paramaters of AQKS and Quantum Hamiltonian}
The AQKS algorithm has several parameters that play a part in its performance as a non-linear kernel transformer.~Notably, the number of episodes, the probability distribution functions (PDF) for choosing $\pmb{\text{A}}$ and ${\bm{b}}$, and the PDFs' parameters are among the tunable parameters.~Unless  otherwise stated, we use a zero-mean Gaussian distribution with a standard deviation of $\sigma_d$ for generating the elements of $\pmb{\text{A}}$, and a uniform distribution for those of ${\bm{b}}$.

In addition to the parameters of the AQKS, the specifications of the adiabatic quantum Hamiltonian and the time evolution process also have the potential to influence the classification accuracy. In our experimentation, we studied the effect of the number of qubits, the annealing time $T$, and the connectivity of the qubits.

\section{Results}
\label{sec:results}
In this section, we report the results of the classification performance for AQKS on both the synthetic dataset and the MNIST dataset.

\subsection{The Synthetic Dataset}
\begin{figure}[h]
    \centering
    \includegraphics[width=0.7\linewidth]{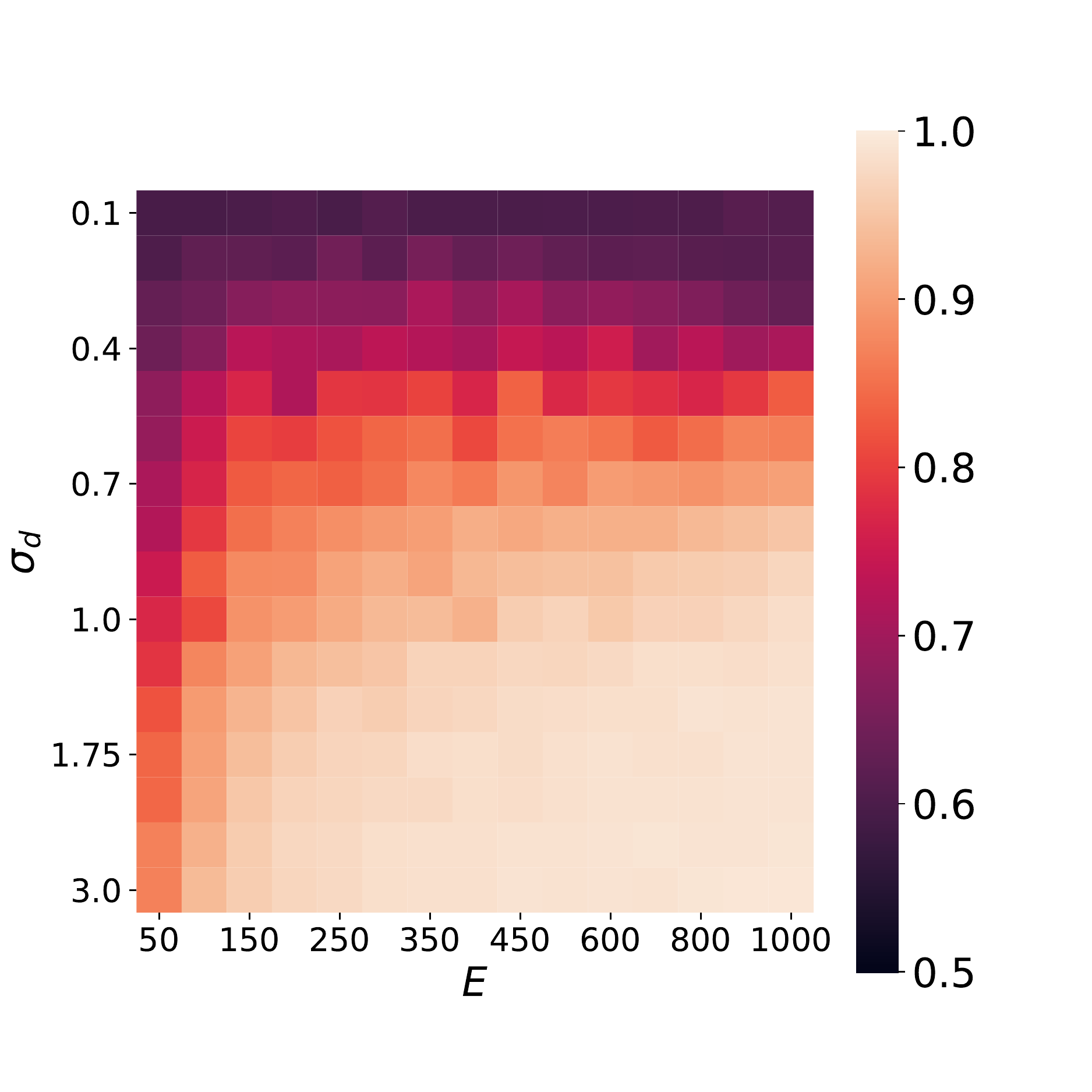}
    \caption{Results of the classification accuracy of AQKS+LSVM with a two-qubit quantum Hamiltonian for the synthetic circles dataset. Here, $\sigma_d$ and $E$ refer to the variance of the normal distribution and the total number of episodes for each setting, respectively. The highest average accuracy that AQKS attains over 10 trials for each setting is $99.4\%$.}
    \label{fig:circular dataset results}
\end{figure}

 Figure~\ref{fig:circular dataset results} is a representation of the classification performance of AQKS using a two-qubit quantum system for the synthetic circles dataset (see Fig.~\ref{fig:circle data}).~The annealing time and duration of the Trotterization time slots are $T=5$ and $\tau=1$, respectively.~Elements of ${\bm{b}}$ are chosen randomly according to a uniform distribution over $[0, 2\pi)$. Applying the LSVM on the quantum-randomized data achieves an average classification accuracy of $99.4$\% averaged over 10 trials of AQKS.

\subsection{The MNIST Dataset}
For the MNIST dataset classification, we first perform  hyperparameter tuning on $\sigma_d$. This is done by fixing the number of qubits to two, the number of episodes to 10,000, and ${\bm{b}}$ to be a zero vector. We run the classification exercise for 3000 images out of the total 13,454 data samples, each time using a different value for $\sigma_d$.~Using $75\%$ of the chosen 3000 images for training and the rest for testing, a value of $\sigma_d=0.01$ yields the best performance.~Similarly, a value of $\sigma_d=0.01$ gives the best result for a four-qubit quantum system.~After finding the optimal value for $\sigma$, we re-run AQKS on all 13,454 samples of data, with $75\%$ used for training and the rest for testing.~The results reported below are for $E = 20000$ and 10 trials of AQKS.

\subsubsection{The Effect of the Number of Qubits}
The mean and standard deviation of the classification accuracy, denoted by $\mu_{\mathrm{c}}$ and $\sigma_{\mathrm{c}}$, respectively, for different numbers of qubits are reported in Table~\ref{table: num qubits}.~For these results, full connectivity between the qubits is assumed, meaning that each qubit interacts with all other qubits in the system. The table shows that quantum randomization improves the classification accuracy of the LSVM. The accuracy improves further as the number of qubits is increased. For instance, increasing the number of qubits from two to four reduces the classification error from $2.3\%$ to $1.6\%$.

\vspace{1.1em}

\begin{table}[h]
    \centering
    \begin{tabular}{cccc}
          &   & Two-qubit system & Four-qubit system \\
         \hline
              Method  & LSVM & AQKS+LSVM  & AQKS+LSVM \\ \hline
         $\mu_\mathrm{c}$ &  $0.951$ & $0.977$  & $0.984$ \\ \hline
         $\sigma_\mathrm{c}$ &  $0.002$ & $0.002$  & $0.002$ \\ \hline
    \end{tabular}
    \caption{Classification accuracy of an LSVM on the randomized features generated by AQKS using a two-qubit and a four-qubit quantum Hamiltonian. The values in the LSVM column show the performance of the  LSVM over the original dataset without any randomization on the input features.}
    \label{table: num qubits}
\end{table}

\subsubsection{The Effect of the Qubits' Connectivity}
\begin{figure}[h]  
    \centering
    \includegraphics[width=0.75 \textwidth]{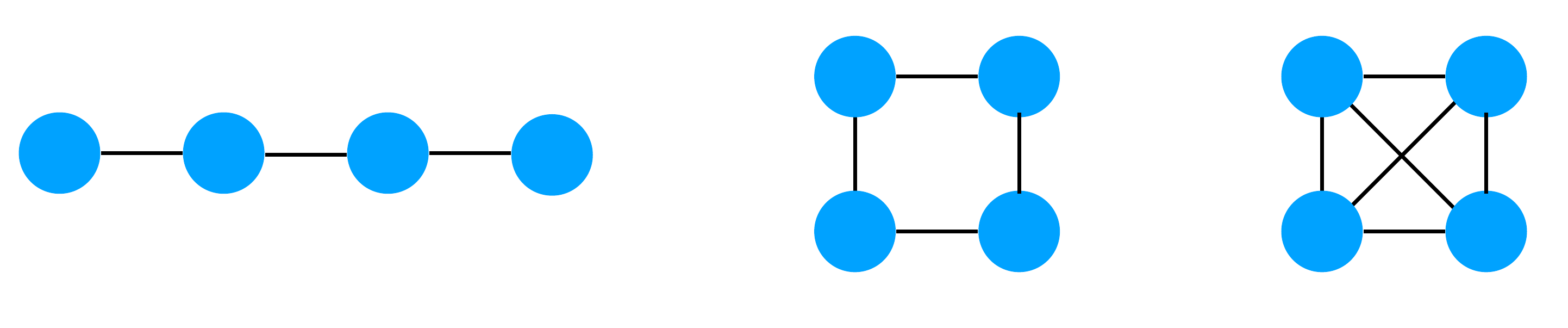}
    \caption{Example connectivity topologies for four qubits. The blue circles represent qubits, and the black lines represent their couplings. We consider (left) a chain of linearly coupled, (middle) nearest-neighbour-coupled, and (right) fully connected qubit architectures.}
    \label{fig: qubit connecitivity}
\end{figure}

We have studied the effect of qubits' connectivity on the classification accuracy of a four-qubit quantum system.~Figure~\ref{fig: qubit connecitivity} shows three example connectivity topologies: linear, square, and complete.~Table~\ref{table: topologies} reports the classification accuracy results for these topologies.

The results imply that for the considered four-qubit quantum Hamiltonian, the topology of the qubits' connectivity does not play a significant role in the classification accuracy for the MNIST dataset.~Regardless, using a different number of qubits (more than four) and episodes, as well as a different choice for the strength of the qubits' coupling, could potentially change this observation.

\vspace{1em}

\begin{table}[h]
    \centering
    \begin{tabular}{cccc}
        Topology &  Linear & Square & Complete\\ \hline
        $\mu_{\mathrm{c}}$ & $0.983$ & $0.984$ & $0.984$ \\ \hline
        $\sigma_{\mathrm{c}}$ & $0.002$ & $0.002$ & $0.002$ \\ \hline
    \end{tabular}
    \caption{Classification accuracy for three connectivity topologies for a four-qubit AQKS. See Fig.~\ref{fig: qubit connecitivity} for examples of connectivity topologies.}
    \label{table: topologies}
\end{table}

\subsubsection{The Effect of Annealing Time}
The effect of annealing time on the performance of a two-qubit AQKS is presented in Table~\ref{table: annealing time}, where the duration of the Trotterization time steps is kept fixed at $\tau=1$.~Results are reported for the classification of only 3000 images out of all 13,454 samples.~Whereas for the considered experiments annealing time does not appear to play a big role, further experimentation is required to be able to arrive at a concrete conclusion. As these experiment are computationally expensive to perform on a classical computer, we have left them for such time as we will be able to employ a quantum device for simulating AQKS. 

\begin{table}[h]
    \centering
    \begin{tabular}{ccc}
        \hline
        $T$ &  $5$ & $50$ \\ \hline
        $\mu_{\mathrm{c}}$ & $0.95$ & $0.96$ \\ \hline
        $\sigma_{\mathrm{c}}$ & $0.01$ & $0.01$ \\ \hline
    \end{tabular}
    \caption{Classification accuracy for different values of the annealing time $T$ for a two-qubit AQKS. Here, $\mu_{\mathrm{c}}$ refers to the mean accuracy and $\sigma_{\mathrm{c}}$ refers to the standard deviation over 10 trials for each setting.}
    \label{table: annealing time}
\end{table}

\section{Discussion}
\label{sec:discussion}
In the case of the synthetic circles dataset (see Fig.~\ref{fig:circle data}), we have deliberately selected a pattern where the two classes of data samples are not linearly separable. Specifically, for the two classes of concentric circles in Fig.~\ref{fig:circular dataset results}, the performance of a linear classifier in two dimensions (i.e.,~a straight line) will not exceed $50\%$. As observed, the accuracy improves further as the number of episodes is increased. In addition, we see that the choice of $\sigma_d$ plays a significant role in the performance of the classification.

For the best set of parameters, namely $E$ and $\sigma_d$, transforming the input feature into a randomized feature space using an adiabatic quantum device improves the performance considerably, bringing it up to $99.4\%$. This is a clear indication that the corresponding quantum kernel \eqref{eq:quantum_kernel_int} has plausible non-linear properties. We highlight again that on both the original and quantum randomized features, an LSVM algorithm is used to perform the classification task. As both the encoding~\eqref{eq:linear_map} and the learning algorithm (LSVM) are linear, it becomes apparent that the observed non-linearity of the quantum kernel is caused by the quantum feature map (i.e.,~the operation we perform using the~adiabatic quantum device).

With respect to the MNIST dataset, using SVMs with linear and RBF kernels, the accuracy of the models trained on this dataset is $95.6\%$ and $99.0\%$, respectively. Compared to the LSVM, the four-qubit AQKS algorithm results in greater accuracy: $98.4\%$. This is another indication that the quantum kernel provides a non-linear property that boosts the performance of the classifier over the performance attained using a linear kernel.

One could still argue that AQKS does not outperform or match an SVM algorithm with a non-linear kernel (e.g.,~RBF). We would like to point out that performing proper hyperparameter tuning for the parameters of our algorithm requires access to a quantum annealer, because simulation of a system comprising four qubits or greater is computationally expensive. In addition to the hyperparameter tuning method, we propose potential modifications to AQKS that could improve its performance in the final section.

\section{Conclusion and Future Work}
\label{sec:conclusion}
In this work, we have introduced a hybrid quantum--classical machine learning algorithm that employs an adiabatic quantum device (a quantum annealer) as an explicit feature map to generate randomized features from input data features. Our algorithm, called ``adiabatic quantum kitchen sinks'',  significantly enhances the performance of a classical linear classifier on the studied binary classification tasks for both the synthetic dataset and the MNIST dataset. Even using the limited-in-size quantum annealers of today~\cite{king2018observation}, our approach can be applied to practical datasets.

In terms of future research, it is worth pointing out that throughout the experiments we performed in our study, we used the same probability distribution functions to generate all the elements of $\pmb{\text{A}}$ and ${\bm{b}}$. We expect that using different types of probability distribution functions for each individual qubit can introduce more-complex forms of non-linearity into the quantum kernel~\cite{CBH18,DUK15}.

One of the advantage of AQKS is that, unlike~\cite{HCT+19,MNK18,SK19}, it does not require constructing a quantum system multiple times in a loop with a classical device. We can, however, modify AQKS to turn it into an adaptive algorithm where we update the parameters of AQKS with respect to the performance of the model, in an iterative fashion. To do so, conside~\eqref{eq:Isingencoding}, and assume that we encode data into the local field parameters ($j$) and that each interacting term ($h^i_{mn}$) represents the adaptive parameters which we intend to update iteratively. Then, for a given machine learning task on a given dataset, we use AQKS to train a model with a generalization error of $F^i$. Having access to $F^i$ and each $h^i_{mn}$, we use a gradient-free optimization algorithm~\cite{NFT19} to update each $h^i_{mn}$ (i.e.,~each adaptive parameter) while reducing the error of the classification accuracy of the model. This iterative process continues until an $F^i$ with a desired threshold has been met or the maximum number of iterations has been reached.

\section{Acknowledgements}
Barry C.~Sanders acknowledges NSERC support. Partial funding for this work was provided by the Mitacs Accelerate program. We thank Marko Bucyk for reviewing and editing the manuscript.


\end{document}